\documentclass{eas}
\usepackage{graphicx}
\begin{document}

\def\CN2{\mbox{$C_N^2 \ $}}
\def\seeFA{\mbox{$\varepsilon_{FA} \ $}}
%%-----------------------------
%%      the top matter
%%-----------------------------
\title{A different {\it glance}  to the site testing above Dome C} 
\author{Elena Masciadri}\address{INAF-Osservatorio Astrofisico di Arcetri, Firenze, Italy; e-mail: masciadri@arcetri.astro.it}
\author{Franck Lascaux}\sameaddress{1}
\author{Jeffrey Stoesz}\sameaddress{1}
\author{Susanna Hagelin}\sameaddress{1}
\author{Kerstin Geissler}\address{European Southern Observatory, Alonso de Cordova 3107, Santiago, Chile}
\begin{abstract}
Due to the recent interest shown by astronomers towards the Antarctic Plateau
as a potential site for large astronomical facilities, we assisted
in the last years to a strengthening of site testing activities in this region, particularly at Dome C. Most of the results collected so far concern meteorologic parameters and optical turbulence measurements based on different principles using different instruments. At present we have several elements indicating that, above the first 20-30 meters, the quality of the optical turbulence above Dome C is better than above whatever other site in the 
world. The challenging question, crucial to know which kind of facilities to 
build on, is to establish how much better the Dome C is than a mid-latitude site. In this contribution we will provide some complementary elements 
and strategies of analysis aiming to answer to this question. We will try to concentrate the attention on critical points, i.e. open questions that still require explanation/attention.
\end{abstract}
\maketitle
%%-----------------------------
%%      your text
%%-----------------------------

\section{Introduction}

\begin{figure}
\center
\includegraphics[width=8cm, bb=30 47 615 603,clip]{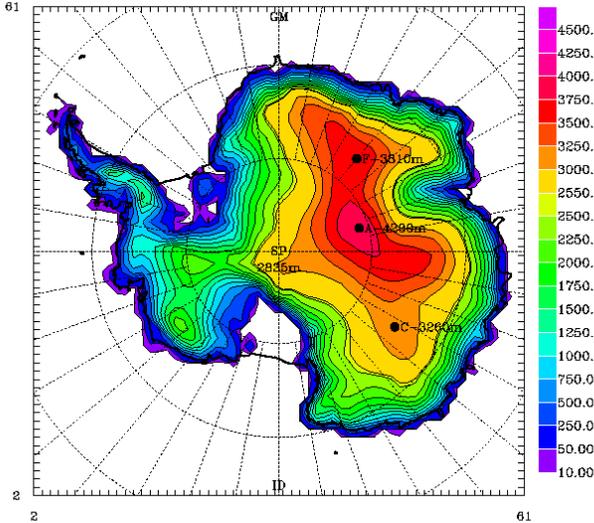}
\caption{Orographic map centred above South Pole and extended on 3000 km $\times$ 3000 km.}
\label{ant}
\end{figure}

Most of the site testing done so far above Dome C has been carried out with measurements taken 
with different instruments and based on different techniques. 
Our approach to the site testing in this context is different. We intend to estimate the optical turbulence ($\CN2$) with the employment of atmospherical models at meso-scale, more precisely we plan to use the Meso-Nh model (Masciadri \& Jabouille 2001, Masciadri, Avila \& Sanchez 2004 , Masciadri \& Egner 2006). This approach provides the possibility to reconstruct 3D $\CN2$ maps (not only vertical profiles) and to forecast the optical turbulence. The flexible scheduling and optimization of scientific observational programs for new class of ground-based telescopes pass necessarily by the forecast of the optical turbulence; this topic is therefore strictly correlated to the success and the survival of the ground-based astronomy of next decades. We recently set-up a research group at the Osservatorio Astrofisico di Arcetri (Florence)\footnote{$http://forot.arcetri.astro.it$} whose aim is to carry out a project (FOROT) centered on the characterization and forecast of the Optical Turbulence. The core project includes the study of two sites: the Mt. Graham, site of the Large Binocular Telescope (LBT) and the Internal Antarctic Plateau (Dome C, South Pole and Dome A). The final goal of our work is to employ this technique to other sites in the world mainly for application to the ELTs. The orographic maps of the Internal Antarctic Plateau, as reconstructed by the Meso-Nh model, is shown in Fig.\ref{ant}. Simulations will be carried out with different horizontal resolution depending on site and scientific goals that we intend to reach.\newline\newline
Why Antarctica ? 
\begin{itemize}
\item Because the whole International Community showed great interest with respect to this site. 
\item Because several site testing campaigns are on-going and this means that a large number of measurements will be soon available. This means a possibility to constrain models outputs.
\item Because we will have the possibility to answer to critical scientific questions among those: (1) ability of meso-scale models in discriminating between sites located on the same plateau but characterized by different turbulence features, (2) characterization of an uncontaminated site such as Dome A, (3) ability of meso-scale models in forecasting the optical turbulence.
\end{itemize}

%\begin{figure}
%\center
%\includegraphics[width=8cm, bb=70 154 510 566,clip ]{f1_masciadri.eps}
%\caption{Orographic map centred above Mt.Graham and extended on a 60 km $\times$ 60 km surface.}
%\label{gra}
%\end{figure}

\section{Challenges at Dome C}
We can summarize the promising potentialities of Dome C for future Astronomy in this Antarctic Continent with the following statement: {\bf at Dome C, above 20-30 m the quality of the atmosphere is better than above whatever other site at mid-latitude.} What does it mean ? Above a good mid-latitude site, the free atmosphere seeing $\seeFA$  (by definition for $h$ $\ge$ 1 km) is typically larger than 0.4 arcsec. Above Dome C, the free atmosphere seeing is $\le$ $0.4$ arcsec and the thickness of the boundary layer is as thin as a few tens of meters ($\sim$ 30 m). An exact estimate of such a thickness is obviously important but we would like to highlight that it has been shown (Ragazzoni et al. 2004) that, using 2 DMs opportunely conjugated in the low part of the atmosphere, it is possible to correct the first $200$ m reaching $15$ arcmin FOV with reasonable good Strehl Ratio. This already does of Dome C an extremely interesting site. The residual turbulence in the free atmosphere, that is not corrected with the system described by Ragazzoni et al. (2004), is critical to assure the predicted performances. For this reason we think that the main priority in a list of studies to be carried out, is to confirm the upper limit for  $\seeFA$ measured so far. As we will explain in the next section, these results need to be confirmed.

\section{The different ``glance''}

Fundamental requirement for our study is to collect as many as possible elements (included measurements) to constrain simulations.  Here we list a few critical points telling us that we still need further studies in the fields of site testing and a different strategic approaches in the organization of site testing campaigns.\newline

{\bf (1)} The vertical profilers: MASS, SSS and Balloons never ran contemporaneously. We can not, at present, know the dispersion between instruments. {\bf (2)} Some instruments, such as the SSS, are prototypes and need careful validation. A comparison made on statistical base between measurements from SSS and another profiler taken as a reference need to be done. Results published so far only concern the comparison between one $\CN2$ profile from SSS and balloon. {\bf (3)} Measurements show discrepancies sometimes. In Agabi et al. (2006), for example, the median isoplanatic angle measured by the DIMM and the Balloons is respectively 2.7 and 4.7 arcsec. This corresponds to a discrepancy of $\sim$ $50$$\%$. What we suspect is that most of the balloons measurements reach the height of 12 km and above this height the turbulence is not measured due to balloons explosion. As a consequence, the isoplanatic angle measured by balloons is larger than the isoplanatic angle measured by a DIMM. 

To supply further elements that might comply with our needs we decided to follow a different approach: to study the analyses provided by the European Center for Medium Weather Forecasts (ECMWF) General Circulation Models. These are vertical profiles of absolute and potential temperature, wind speed and direction extracted in the grid point (75$^{\circ}$ S, 123$^{\circ}$ E), horizontal resolution 0.5$^{\circ}$, extended up to 0.1 hPa, 60 vertical levels with resolution decreasing with height. Data have been treated statistically on the 2003-2004 period at heights {\it h} $\ge$ 30 m. The scientific goals of this study are: (1) to provide a yearly meteorologic parameters analysis, (2) to calculate the probability to trigger optical turbulence in different slabs of atmosphere and in different periods of the year and (3) to estimate the quality of ECMWF analyses in perspective to use these data to initialise Meso-Nh.

\begin{figure}
\centering
\includegraphics[width=5cm,angle=-90]{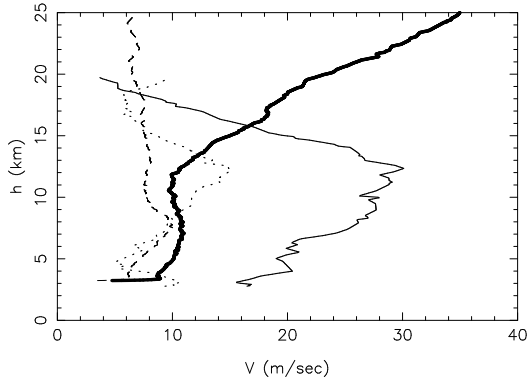}
\caption{Comparison between the median wind speed profile estimated during the winter time (full bold line) and summer time (dashed line line) (2003) above Dome C and the median wind speed profile estimated above the San Pedro M\'artir Observatory in summer (dotted line) and winter (full thin line) time. San Pedro M\'artir is taken as representative of a mid-latitude site.}
\label{spm_domec}
\end{figure}

\begin{figure}
\centering
\includegraphics[width=3.7cm,angle=90,clip]{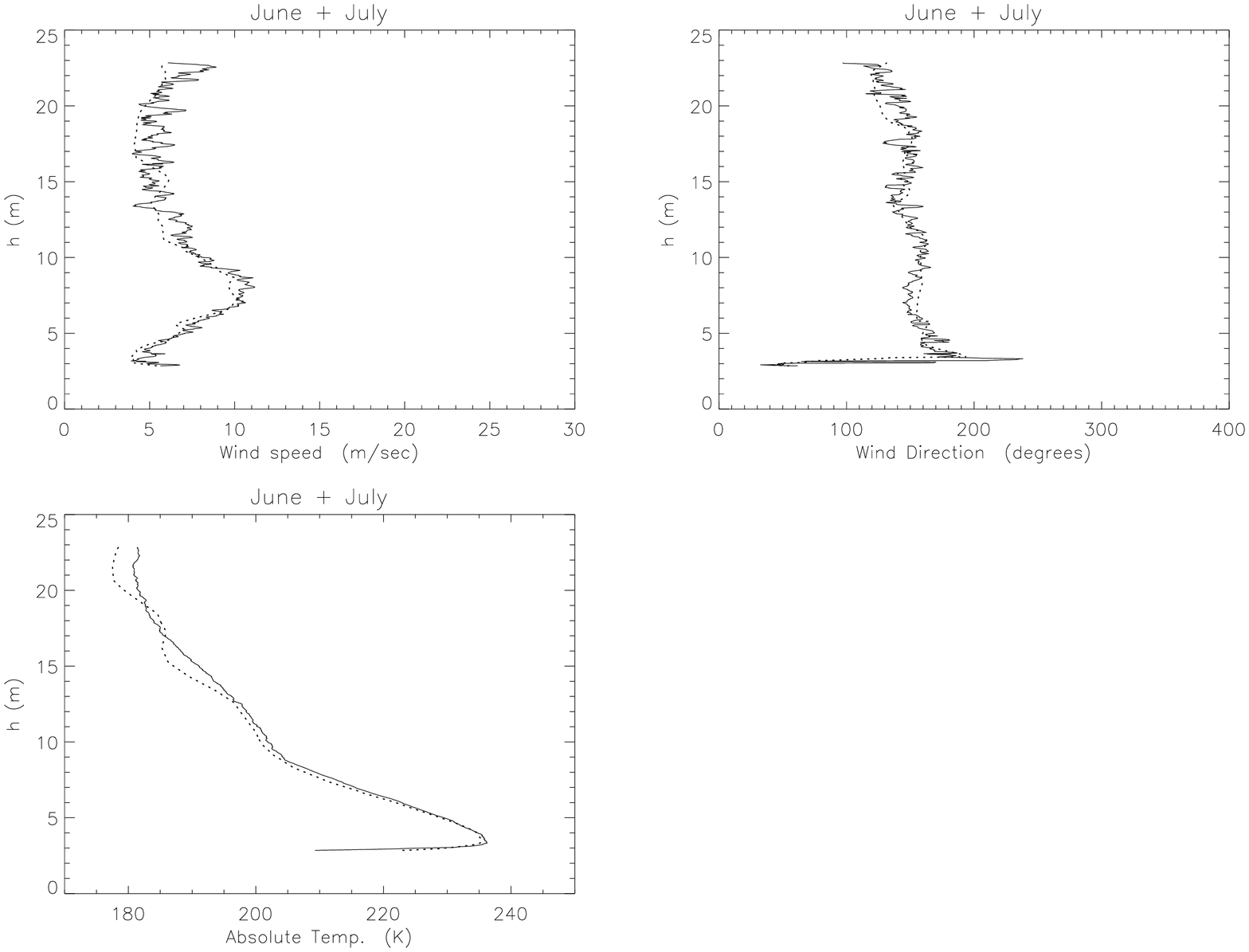}
\qquad
\includegraphics[width=4.6cm,clip]{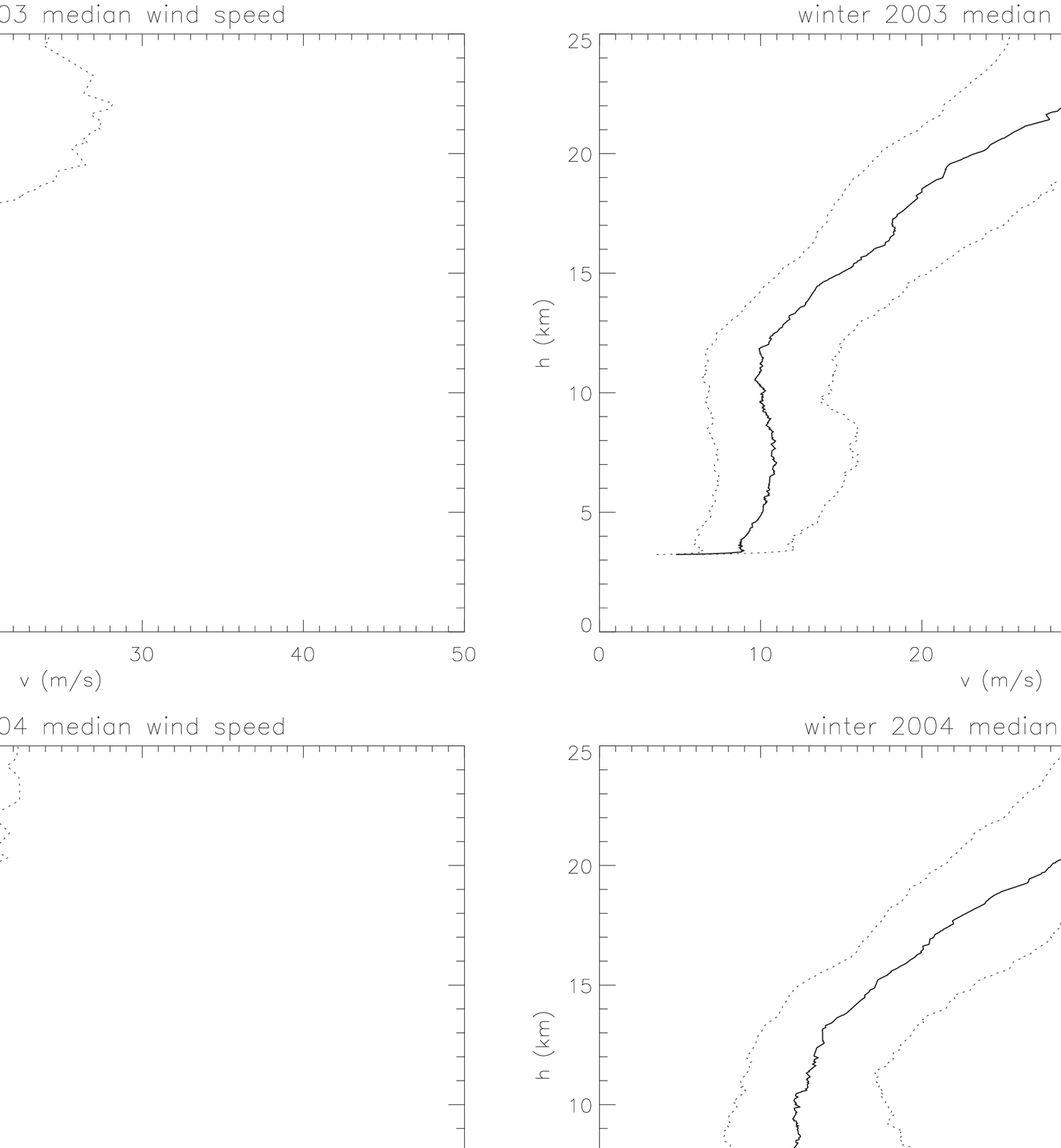}
\caption{ECMWF analyeses estimates. Left: median wind speed vertical profile in winter time (June and July) above South Pole (dotted line). Right: median wind speed vertical profile in winter time above Dome C (dotted lines are first and third quartiles). See Geissler \& Masciadri (2006) for further details.}
\label{sp_dc}
\end{figure}

Analyses reliability has been proved in Geissler \& Masciadri (2006). In that paper the main results of this study are extensively described.  Figure \ref{spm_domec} shows the median wind speed in summer and winter time above 2 sites: Dome C and San Pedro M\'artir. The latter has been selected as representative of a mid-latitude site. It is possible to observe that above Dome C the wind speed is particularly weak in summer time on the whole troposphere; a peak appears at around 5 km above the ground (tropopause). In winter time the wind speed increases monotonically above 10 km above the ground reaching 30 m/sec above 20 km. Such a strong wind speed should be considered with attention for at least 2 reasons. Firstly it might trigger dynamic instabilities in the high free atmosphere, secondly it might affect the wavefront coherence time $\tau_{0}$. This parameter indeed scales as $C_{N}^{2}(h)\cdot \left\vert V(h)\right\vert ^{5/3}$, i.e. is particularly sensitive to large wind speed values.\newline\newline
Fig.\ref{sp_dc} shows the median wind speed vertical profile in winter time (June and July) above South Pole and the median wind speed vertical profile in winter time above Dome C. Both results are retrieved by statistical estimates of ECMWF analyses. It is interesting to note that the wind speed at high altitudes is much weaker at South Pole than at Dome C. \newline\newline
How we can connect meteorological and astroclimatic parameters ? The study of the Richardson number $R_{i}$$=$$R_{i}$(${\partial \theta }/{\partial h}$,${\partial V}/{\partial h}$), parameter indicating the probability to trigger turbulence, shows (Geissler \& Masciadri 2006, Fig.14) a monotonic increasing of the turbulence activity above 15 km from summer towards and across the winter time. One would expect therefore a more intense turbulence activity at high altitudes during the winter than during the summer time. The trend identified by the Richardson maps is confirmed by measurements of $\theta_{0}$ obtained with the same instrument in summer and winter time. $\theta_{0}$ is much larger in summer time (6.8 arcsec - Aristidi et al. (2005)) than in winter time (2.7 arcsec - Agabi et al. (2006)). This means that turbulence at high altitudes is weaker in summer than in winter time. We should therefore increase the statistic of measurements in August, September and October to better characterize this phenomenon.

A further result related to the seeing in the free atmosphere ($\varepsilon_{FA}$) requiring attention and careful verification is the following. We can retrieve from the Fig.5 (Agabi et al. 2006) that seeing measured by a DIMM placed at 20 m above the ground grows in a monotonic way reaching values greater than 1 arcsec in August. Where this turbulence come from if the seeing above 30 m has been estimated not greater than 0.4 arcsec during the winter time ? One could think that this is the contribution associated to the thin layer between 20 and 30 m. If we look at the typical $\CN2$ values at 20-30 meters (10$^{-14}$ $\le$ $\CN2$ $\le$ 5 $\times$ 10$^{-14}$) (Fig.3, Agabi et al. 2006), we calculate the correspondent seeing in the 20-30 m slab and we finally subtract\footnote{The seeing subtraction is done following the $\left( \sum \left( \varepsilon _{i}\right) ^{5/3}\right) ^{3/5}$ rule.} the result from the seeing reported in Fig.5, we find a seeing in the range 0.45 - 0.92 arcsec (at h $\ge$ 30 m). We therefore obtain a seeing above 30 m that is greater than the 0.4 arcsec indicated by preliminary results. This exhorts us in pushing on further studies of seeing in this region of the atmosphere.

\section{Conclusion}
We summarize the main points discussed in this contribution:
{\bf (1)} We indicated an alternative method (ECMWF analyses) to check measurements, to constrain simulations and characterize climatology above Dome C.
{\bf (2)} Further optical turbulence measurements are necessary to carry out all scientific programs related to site testing. 
However it would be suitable that future site testing campaigns are planned conjointly with all teams working on site testing
taking into account requirements of all scientific programs. This is the only way to maximize the scientific outputs of 
activities supported by an international logistic (IPEV and PNRA) and funded by national and international structures such as the EC. {\bf (3)} We recommend to run the SSS in contemporary way to the DIMM (N$^{\circ}$ 1) adapted to isoplanatic angle measurements placed at the same height than SSS (8 m) and to the DIMM (N$^{\circ}$ 2) placed at 20 m. It would be fundamental to collect a large statistic employing this instruments configuration. In this way we will have 3 independent ways to measure the free atmosphere and it will be able to cross-correlate results in this particular critical region of the atmosphere. This approach is particularly important for the FOROT activities and also, as explained in Section 2, to design the future of Astronomy at Dome C.

\section{Acknowledgments}

This work has been funded by the Marie Curie Excellence Grant (FOROT) - MEXT-CT-2005-023878

%%-----------------------------
%%      your bibliography
%%-----------------------------

\end{document}